%% file: main18ibbAA.tex
\documentclass{aa}

\usepackage{graphicx}
\usepackage{txfonts}
\usepackage{hyperref}
\usepackage{booktabs}
\usepackage{bm}
\usepackage[normalem]{ulem}
\usepackage[dvipsnames]{xcolor}
\usepackage{nicefrac}
\usepackage{xspace}

\usepackage{mathtools}
\usepackage{hyperref}
\hypersetup{
  colorlinks,
  citecolor=blue,
  linkcolor=blue,
  urlcolor=blue}
\usepackage{amsmath}
\usepackage{yhmath}
\usepackage{multirow}
\usepackage{array}

\newcommand{\Msun}{$\mathrm{M}_\odot$\xspace}
\newcommand{\Rsun}{$\mathrm{R}_\odot$}

\newcommand{\appropto}{\mathrel{\vcenter{
  \offinterlineskip\halign{\hfil$##$\cr
    \propto\cr\noalign{\kern2pt}\sim\cr\noalign{\kern-2pt}}}}}

\begin{document}

\title{Primary and secondary source of energy in\\the superluminous supernova 2018ibb
\thanks{The data computed and analysed for the current study are available via the link
\href{https://zenodo.org/doi/10.5281/zenodo.10473678}{https://doi.org/10.5281/zenodo.10473679}.
}
}
\titlerunning{SLSN 2018ibb as PISN}


\author{
    Alexandra Kozyreva\inst{1},
    Luke Shingles\inst{2},
    Petr Baklanov\inst{3,4,5}\and
    Alexey Mironov\inst{6},
    Fabian R.\,N.\,Schneider\inst{1,7}
}
\authorrunning{A. Kozyreva et al.}
\institute{
    Heidelberger Institut f{\"u}r Theoretische Studien,
    Schloss-Wolfsbrunnenweg 35, D-69118 Heidelberg, Germany\\
    \email{sasha.kozyreva@gmail.com}
    \and
    GSI Helmholtzzentrum f\"ur Schwerionenforschung, Planckstra\ss e 1, 64291 Darmstadt, Germany
    \and
    National Research Center, Kurchatov Institute, pl. Kurchatova 1, Moscow 123182, Russia 
    \and
    Lebedev Physical Institute, Russian Academy of Sciences, 53 Leninsky Avenue, Moscow 119991, Russia
    \and
    Space Research Institute (IKI), Profsoyuznaya 84/32, Moscow 117997, Russia
    \and
    M.V. Lomonosov Moscow State University, Sternberg Astronomical Institute, 119234, Moscow, Russia
    \and
    Astronomisches Rechen-Institut, Zentrum f\"ur Astronomie der Universit\"at Heidelberg, M\"onchhofstr. 12-14, D-69120 Heidelberg, Germany
    }

\date{Received; accepted }

\abstract{
We examine the pair-instability origin of superluminous supernova 2018ibb. As the base model, we use a non-rotating stellar model with an initial mass of 250~\Msun at about 1/15 solar metallicity. We consider three versions of the model as input for radiative transfer simulations done with the \textsc{stella} and \textsc{artis} codes: with 25~\Msun{} of {}$^{56}$Ni, 34~\Msun{} of {}$^{56}$Ni, and a chemically mixed case with 34~\Msun{} of {}$^{56}$Ni. We present light curves and spectra in comparison to the observed data of SN\,2018ibb, and conclude that the pair-instability supernova model with 34~\Msun{} of {}$^{56}$Ni explains broad-band light curves reasonably well between $-100$ and 250~days around the peak. Our synthetic spectra have many similarities with the observed spectra. The luminosity excess in the light curves and the blue-flux excess in the spectra can be explained by an additional energy source, which may be interaction of the SN ejecta with circumstellar matter. We discuss possible mechanisms of the origin of the circumstellar matter being ejected in the decades before the pair-instability explosion.
}

\keywords{supernovae --- white dwarf --- giant star--- common envelope 
--- stellar evolution --- radiative transfer}

\maketitle

\input{intro}
\input{method}
\input{results}
\input{conclusions}

\section*{Acknowledgments}

We thank Steve Schulze, Pavel Abolmasov, Marat Potashov, Elena Sorokina, and Sergei Blinnikov for helpful discussions.

AK acknowledges support by Alexander von Humboldt Stiftung.
LJS acknowledges support by the European Research Council (ERC) under the European Union’s Horizon 2020 research and innovation program (ERC Advanced Grant KILONOVA No. 885281).
LJS acknowledges support by Deutsche Forschungsgemeinschaft (DFG, German Research Foundation) - Project-ID 279384907 - SFB 1245 and MA 4248/3-1.
PB is sponsored by grant  RSF\,23-12-00220 in his work on the \verb|STELLA| code development.

This project began before February 2022.


\bibliography{references}{}
\bibliographystyle{aa}

\end{document}

%% file: intro.tex

\section[Introduction]{Introduction}
\label{sect:intro}

The superluminous supernova (SLSN, SN) 2018ibb is a unique supernova (SN) event that was discovered long before its peak
\citep{2018TNSTR1722....1T}.
It was classified as a hydrogen and helium free, i.e., SLSN Type\,I, when the first spectral data were collected around its maximum \citep{2018TNSCR2184....1P}\footnote{The first classification was done by \citet{2018TNSCR1877....1F}, however, they misclassified SN\,2018ibb as SN\,Ia.}. SN\,2018ibb has an exquisite
follow-up campaign and exclusively good data coverage, starting from 100~days before peak magnitude, 109~days in the Gaia/$G$--band in the observer's frame \citep{2024A&A...683A.223S}.
The absolute peak magnitude of $-21.5$~mags and the long rise of 100~days to the peak indicate a high mass of radioactive nickel {}$^{56}$Ni, about 30~\Msun{}, and a long diffusion time, which implies a high ejecta mass. 
From a theoretical point of view, it is well known that stars with initial hydrogenic mass range between 140\,--\,260~\Msun{} undergo pair instability (PI) and explode as pair-instability supernovae \citep[PISN,][]{1967PhRvL..18..379B,1967ApJ...148..803R}.
Depending on the mass of a carbon-oxygen core formed by the PI episode,
a broad range of a total yield of radioactive {}$^{56}$Ni are produced, from zero to up to 60~\Msun{} \citep{2002ApJ...567..532H,2016MNRAS.456.1320T}. 
Therefore, the resulting SN may be either sub- or superluminous
\citep[$-14$\,mag to $-22$\,mag in $V$; ][]{2011ApJ...734..102K} with broad light curves which rise to the peak during 100 to 150~days.

In the present study, we examine and confirm a possible PISN origin of SN 2018ibb\footnote{Another study exploring different PISN models in application to SN\,2018ibb  was submitted during the reviewing stage of the present paper \citet{2024arXiv240416570N}. Based on the derived photospheric parameters, the authors rule out the PISN origin of SN\,2018ibb.}. 
We specifically chose SN\,2018ibb because it passed a number of tests showing that no other mechanisms, such as magnetar or pure interaction can explain its origin. For instance, SN\,2018ibb being observed for more than 1000~days does not show a signature of different slope at late time as expected from magnetar-powered SNe \citep{2017ApJ...835..177M}. SN\,2018 is the first most reliable candidate for PISNe. Other SLSNe Type\,Ic are supposedly powered by a magnetar or interaction \citep{2013Natur.502..346N,2015MNRAS.452.1567C}.

Details of our progenitor and explosion models and the 
techniques to model light curves and spectra are given in 
Section~\ref{sect:method}. In Section~\ref{sect:res}, we present 
synthetic observables for our models, describe comparison to observations, and discuss the primary and secondary sources of energy in SN\,2018ibb. We summarise our study in Section~\ref{sect:conclusions}.

%% file: method.tex

\section[Progenitor model and modeling of light curves and spectra]
{Progenitor model and modeling of light curves and spectra}
\label{sect:method}

\subsection[The choice for the evolutionary model]
{Explosion models}
\label{subsect:method1}


To support high luminosity of SN\,2018ibb of $\log L \sim 44.25$\,erg\,s$^{\,-1}$ about 36~\Msun{} of radioactive nickel $^{56}$Ni is required assuming the nickel-powering mechanism \citep{2016MNRAS.459L..21K}. According to \citet{2002ApJ...567..532H}, this amount of $^{56}$Ni can be produced by a PI explosion of a star with the He-core of 126~\Msun{}, in turn by an originally hydrogenic star with initial mass of about 250~\Msun{} \citep[see Eq.\,1 in][]{2002ApJ...567..532H}. Therefore, we focus our modelling on a 
very massive non-rotating star model, P250, with M$_{\rm
ZAMS}$ = 250~\Msun{} at a metallicity of \textit{Z}=0.001 \citep[$Z\cong 0.07\, Z_\odot${}, where $Z_\odot=0.014${}, ][]{2005ASPC..336...25A}. 
Our choice of metallicity of 0.001 is not precisely consistent with the claimed metallicity of the host of SN\,2018ibb, $0.25\,Z_\odot${} \citep{2024A&A...683A.223S}. We note though, that the host metallicity, given with some degree of uncertainty, is an averaged metallicity, based on the measurements of many HII regions, and it could not be specified for the exact SN explosion site. In any case, both the host metallicity and the stellar model metallicity are sub-solar.
The stellar model P250 was evolved from the zero age main sequence (ZAMS) until the start of the PI phase with the stellar evolution code \textsc{genec} \citep[][]{2012A&A...537A.146E,2013MNRAS.433.1114Y}.
The radius of the model is 2~\Rsun{} and the mass is 127~\Msun{}
prior to the PI. The star consists mainly of carbon and oxygen and has a shallow helium layer at the
surface, with a total helium mass of 2.6~\Msun{}.
PI explosion of the original model P250 is carried out with the \textsc{flash} code
\citep{2000ApJS..131..273F,2009arXiv0903.4875D,2013ApJ...776..129C,2015ApJ...799...18C}
while mapping the hydrodynamical and chemical profiles from
\textsc{genec} into \textsc{flash}. All details about the \textsc{flash}
simulations are published in \citet{2017ApJ...846..100G} and \citet{2017MNRAS.464.2854K}.
We use the result of \textsc{flash} simulations, in which 33.8~\Msun{} of radioactive nickel {}$^{56}$Ni{} are produced during the explosive oxygen and silicon burning. The explosion energy of this model carried by the ejecta in the form of thermal and kinetic energy is 84~foe, where 1~foe $=10^{\,51}$\,erg. We name this model as P250Ni34. The model P250Ni34mix is the same as P250Ni34 which is 
chemically homogeneously mixed, i.e., all yields are kept as in the original model P250Ni34, but all species are distributed uniformly through the ejecta without changing other hydrodynamical profiles, similar to \citet{2019MNRAS.484.3451M}. We also built another model, P250Ni25, based on the model P250Ni34 with a reduced amount of {}$^{56}$Ni{}, 24.6~\Msun{},
while scaling the profile of {}$^{56}$Ni{} with a factor of 2/3 to
check whether the smaller yield of radioactive nickel {}$^{56}$Ni{} might power the light
curve of SN 2018ibb. The missing fraction of {}$^{56}$Ni{} was replaced with silicon.
A lower yield of {}$^{56}$Ni{}, 24.6~\Msun{}, was also obtained in one of the 
test \textsc{flash} simulations (M.\,Gilmer, private communication).

\subsection[Light curve modelling]{Light curve modelling}
\label{subsect:method2}

We follow the supernova explosion and model the light curves
with the one--dimensional multigroup radiation hydrodynamics code
\textsc{stella}
\citep[][]{1998ApJ...496..454B,2000ApJ...532.1132B,2006AandA...453..229B}. 
The description of \textsc{stella}, including, particularly, details about opacities, and comparison to other radiative transfer codes are presented in \citep{2022A&A...668A.163B}.
We mapped the models into \textsc{stella} when the shock was close to the edge of the star, i.e. shock breakout. In the present simulations, we use 100 frequency bins and spacial grid of 231 zones.

\subsection[Spectral synthesis]{Spectral synthesis}\label{subsect:spectra}


We use the Monte Carlo radiative transfer code \textsc{artis} \citep{2009MNRAS.398.1809K} in full non-LTE\footnote{LTE --- local thermodynamic equilibrium.} mode \citep{2020MNRAS.492.2029S} to compute spectral time series for the models in our study.
The full non-LTE treatment includes deposition by non-thermal electrons, non-LTE ionisation balance and level populations, detailed bound-free photoionisation estimators using full packet trajectories
\citep[equation 44 of][]{2003A&A...403..261L} for all photoionisation transitions, and an non-LTE binned radiation field model for excitation transitions. 
Forbidden transitions are included in the dataset based on the atomic data compilation of \textsc{cmfgen}\footnote{Available at
\url{http://kookaburra.phyast.pitt.edu/hillier/web/CMFGEN.htm}}
\citep{1990A&A...231..116H,1998ApJ...496..407H} with typically the first five ionisation stages as detailed in \citet{2020MNRAS.492.2029S}.

As an input model for our \textsc{artis} simulations, we use the spherically-symmetric density and abundance snapshot from \textsc{stella} at 10\,d after the explosion, when the ejecta approach homologous expansion\footnote{Note, though, that the effect of Ni--bubble develop on a timescale of 100~days, however, there is no significant contribution to the observables. See discussion in \citet{2017MNRAS.464.2854K}.}.

We simulate the evolution of the energy flows and radiation field by propagating $5\times10^{\,8}$\, Monte Carlo quanta for 150 logarithmically-spaced time steps from 150 to 850\,d past explosion. The number of interactions per packet was larger than one for a timespan exceeding 600~days after the explosion. Once the Monte Carlo quanta escape from the simulation domain they are binned in time and on a logarithmic wavelength grid spanning 1,000 bins from 375\,\AA\ to 30,000\,\AA\ to obtain the spectral time series\footnote{For the Ni34 model, computational cost was 367 hours on 960 cores (352~k core hours). For Ni25, the cost was 422 hours on 960 cores (405~k~core hours).}.

In \textsc{STELLA} simulations, the source of radioactivity is presented only by radioactive nickel {}$^{56}$Ni, while \textsc{artis} includes also radioactivity from {}$^{57}$Ni{}. Therefore, we artificially change chemical composition of the input models to consider {}$^{57}$Ni{} in \textsc{artis} simulations. 
According to the set of zero-metallicity helium PISN models by
\citet{2002ApJ...567..532H}, the total yield of $^{57}$Ni{} is about 1/50 of the {}$^{56}$Ni{} yield. Therefore, we set 0.5~\Msun{} and 0.7~\Msun{} of {}$^{57}$Ni{} in the chemical profiles mapped into \textsc{artis} for the models P250Ni25, and P250Ni34 and P250Ni34mix, respectively. However, the {}$^{57}$Ni{} yield differs for very massive stars at non-zero metallicity, which is the case of our P250 with the initial metallicity of 0.001. The yield of {}$^{57}$Ni{} might be higher, up to 10\,\%{}, relative to {}$^{56}$Ni{} yield, since neutron excess is higher for higher metallicity \citep{2002ApJ...565..385U,2014A&A...566A.146K,2018ApJ...857..111T}.

%% file: results.tex

\section[Results]{Results and Discussion}
\label{sect:res}

In the following, firstly, we analyse the synthetic light curves in the context of bolometric and broad-band light curves of SN\,2018ibb, secondly, we present the result of spectral synthesis and compare our models to the observed spectra of SN\,2018ibb. Based on this analysis we draw conclusion about the circumstances of the explosion of SN\,2018ibb progenitor.

\subsection[Bolometric and broad bands light curves]{Bolometric and broad bands light curves}
\label{subsect:broadbands}

\begin{figure}
\centering
\hspace{-5mm}\includegraphics[width=0.5\textwidth]{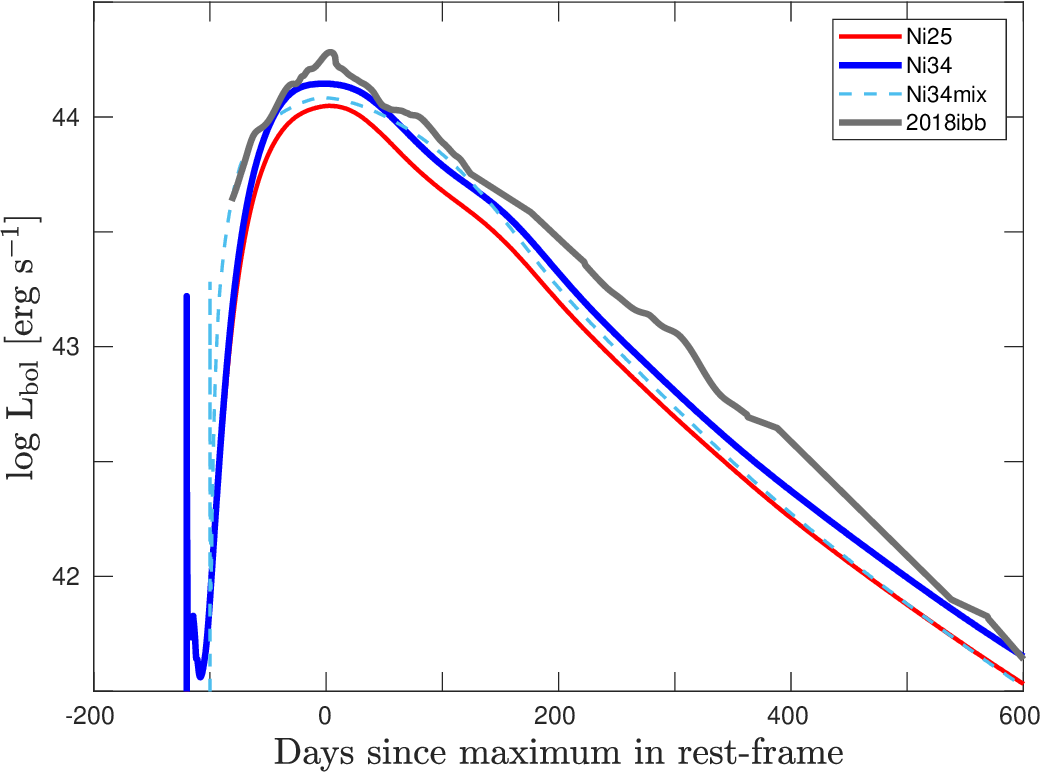}
\caption{Bolometric light curves for the models P250Ni25 (red curve, label ``Ni25''), P250Ni34 (blue curve, label ``Ni34''), and P250Ni34mix (cyan curve, label ``Ni34mix''), and the bolometric light curve for SLSN\,2018ibb (grey curve, label ``2018ibb'') taken from \citet{2024A&A...683A.223S}.  Time ``0'' corresponds to the peak in \emph{g/r}-band magnitude, similar to $t_\mathrm{max}$ in \citet{2024A&A...683A.223S}.}
\label{figure:lbol}
\end{figure}

\begin{figure*}
\centering
\hspace{-5mm}\includegraphics[width=\textwidth]{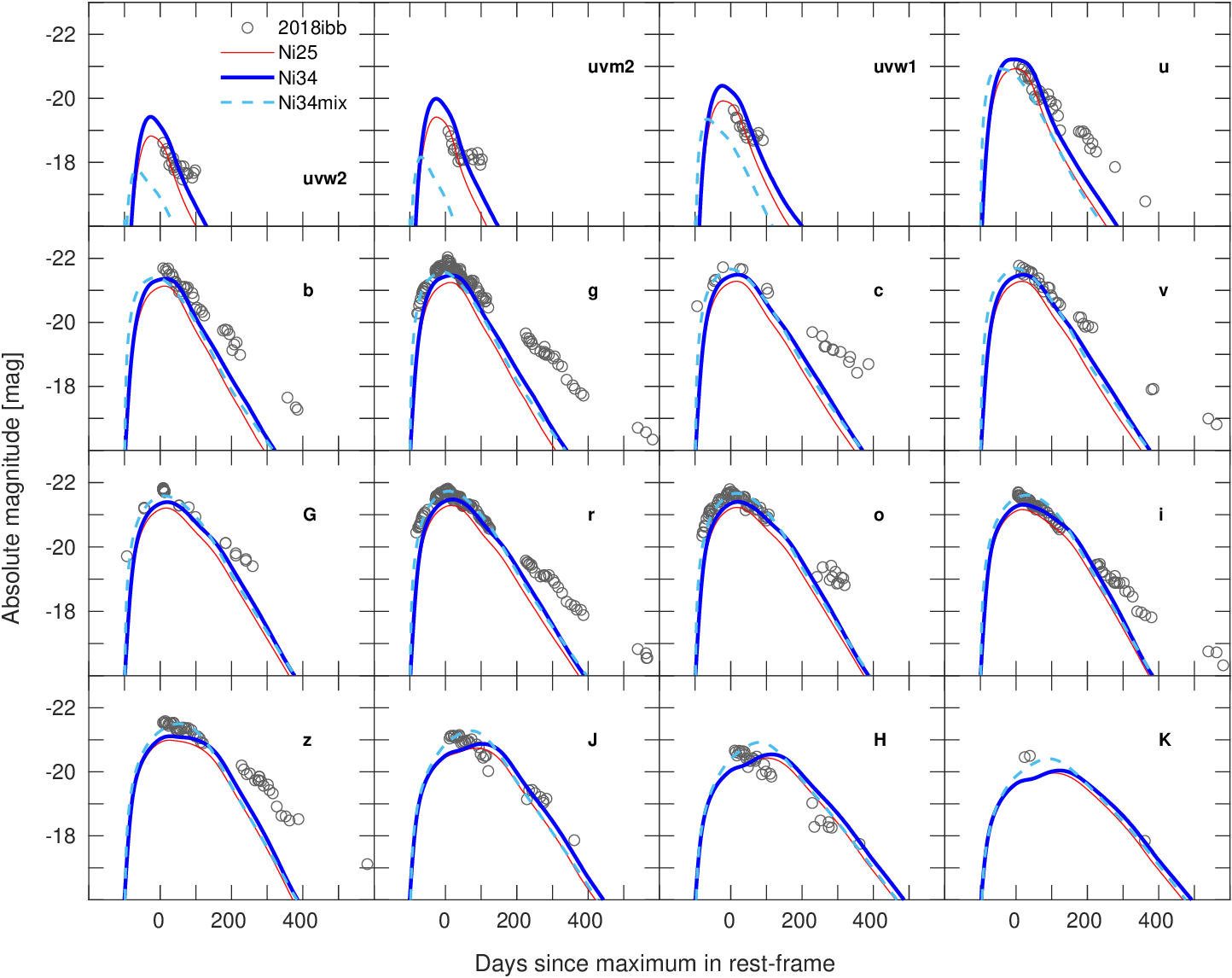}
\caption{Broad-band magnitudes for three models in the study in comparison to those for SN 2018ibb \citep[grey circles, ][]{2024A&A...683A.223S}. Red, blue and cyan curves represent P250Ni25, P250Ni34 and P250Ni34mix models, respectively.
}
\label{figure:bands}
\end{figure*}

In Figure~\ref{figure:lbol}, we present bolometric light curves for the models P250Ni25, P250Ni34, and P250Ni34mix in the rest frame\footnote{We note that the present light curve for the model P250Ni34 differs from those published in \citet{2017MNRAS.464.2854K} and \citet{2024A&A...683A.223S}, because we fixed a numerical issue caused by high velocity of the ejecta.}. Time $t=0$ is set similar to $t_\mathrm{max}=0$ (MJD=58455) in \citet{2024A&A...683A.223S} for consistency and corresponds to the SN\,2018ibb peak in $g/r-$bands. Our synthetic light curves are true bolometric, and correspond to integrated flux between 1\,\AA{} and 50,000\,\AA{}.
Bolometric light curve for SN\,2018ibb (shown as a grey curve in Figure~\ref{figure:lbol}) is taken from \citet{2024A&A...683A.223S}. In their study, bolometric light curve is constructed differently at different epochs. For instance,  luminosity between day~0 and day~100 is integrated over wavelength range 1800\,\AA{}--\,14,300\,\AA{}, because there is a complete photometric coverage in this time window. However, the part of the bolometric light curve before the peak is based on the observed flux in ZTF $g$- and $r$-bands and correction estimated at the first epoch with good coverage \citep[see all details in Section~4.2.2 in][]{2024A&A...683A.223S}.
In Figure~\ref{figure:bands}, we present our resulting light curves 
in broad bands $uvw2$, $uvm2$, $uvw1$, $u$, $b$, $g$, $c$, $v$, $G$, $r$, $o$, $i$, $z$, $J$, $H$, $K$ in the rest-frame and observed broad-band light curves for SN\,2018ibb \citep{2024A&A...683A.223S}. The rise time for the models P250Ni25 and P250Ni34 is 120~days, and for the model P250Ni34mix is 100~days in the rest-frame. 

For the available data there is generally good agreement of our model P250Ni34 starting from the non-detection epoch until day~250 after the peak, i.e., over 450~days. Therefore, it is clear that the primary source of energy powering the light curve of SN\,2018ibb is radioactive nickel $^{56}$Ni produced in the PI explosion. However, there is some difference between the observed and synthetic light curves, which we discuss below.
The model P250Ni25 clearly underestimates the flux in many broad bands, even though it still agrees reasonably well with SN 2018ibb in $uvw2$, $uvm2$, $uvw1$, and $u$-bands.
The uniform mixing in the model P250Ni34mix helps to reduce rise time to 100~days in the rest-frame. In this model, radioactive {}$^{56}$Ni{} is distributed up to the edge while being mixed, therefore, diffusing high energy photons reach the emitting front faster. Consequently, P250Ni34mix makes a perfect fit to the rising part of the SN\,2018ibb bolometric light curve. However, the photons then easily escape from the ejecta, lowering the flux in $uvw2$, $uvm2$, and $uvw1$ bands and weakening the suitability of the model P250Ni34mix.
Furthermore, bolometric luminosity at peak is also lower, because the amount of radioactive energy contained in the inner region is reduced due to the lower mass fraction of {}$^{56}$Ni{} compared to the unmixed case P250Ni34, although the total mass of {}$^{56}$Ni{} is the same. We note that light curves in $J$ and $H$ bands systematically deviate from the observed light curves. Likely, the list of line transitions implemented in \textsc{stella} is not exhaustive enough to make a solid prediction for flux in these bands, although good agreement to more sophisticated spectral synthesis codes show suitability of the \textsc{stella} simulations for these wavelengths in application to SNe\,Ia \citep[see Fig. 8 in][]{2022A&A...668A.163B}.

Our models deviate noticeably from observations later than day~200 after the peak. There are a few episodes, which we call ``bumps'' later in the paper, where the observed light curve clearly deviates from the modelled light curve. Likely, another, secondary, energy source contributes to the light curve at later time, which we discuss below.


\subsection[Spectral comparison to SN\,2018ibb]{Spectral comparison to SN\,2018ibb}
\label{subsect:spectr}

\begin{figure*}
\centering
\hspace{-5mm}\includegraphics[width=\textwidth]{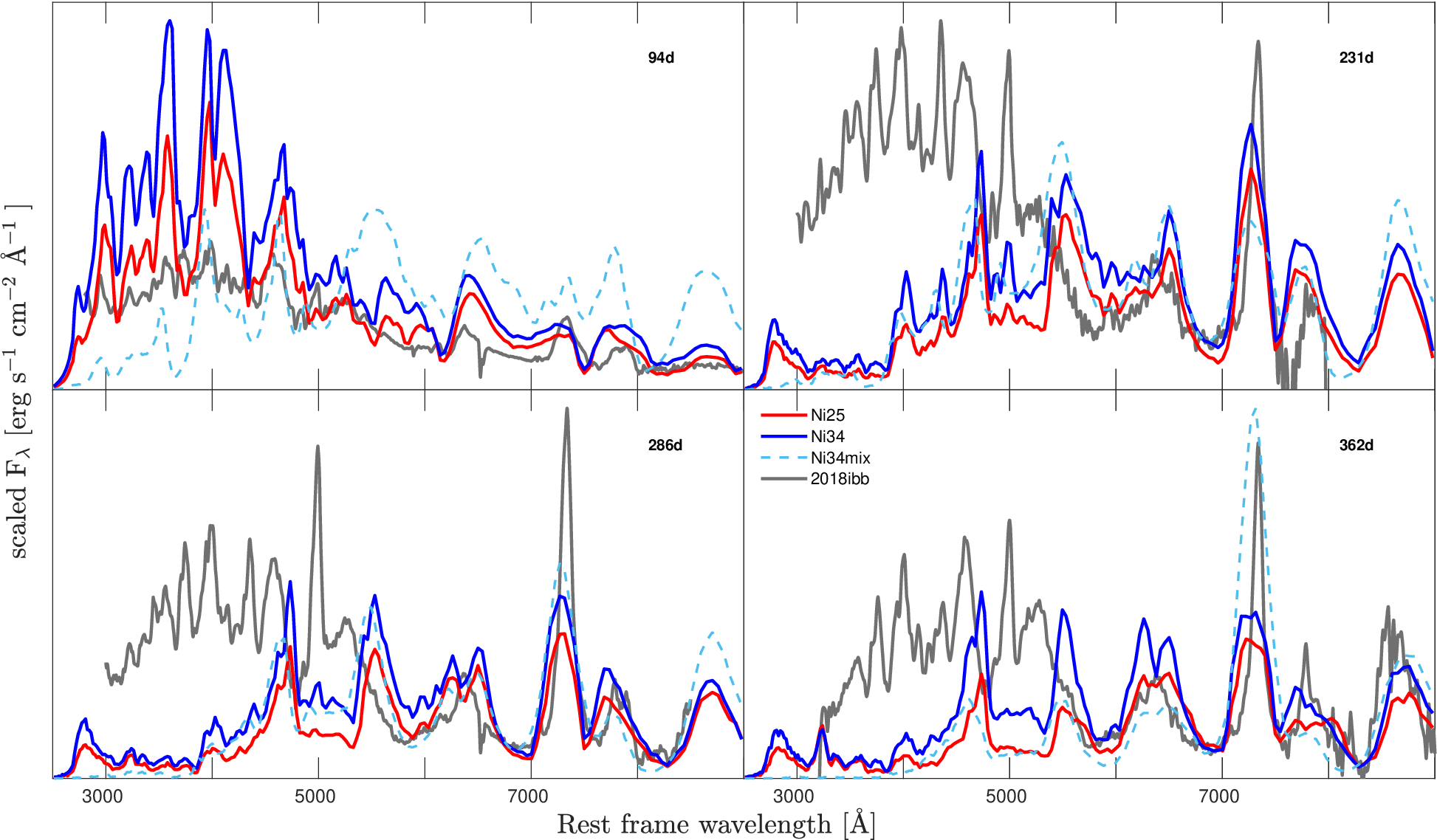}
\caption{Spectral series of synthetic spectra with superposed observational data.
Red, blue and cyan curves represent P250Ni25, P250Ni34 and P250Ni34mix models, respectively. The observed spectra (shown in grey) presented in the plot are different to those published in \citet{2024A&A...683A.223S}, because we plotted spectra corrected for galactic extinction and the host (S.\,Schulze, private communication). 
Days are since the $g/r$-maximum in the rest-frame. 
Spectra are calibrated to the maximum flux values in each subplot. 
Rise time of 120 days and 100~days has been assumed for P250Ni34 and P250Ni25, and P250Ni34mix models.}
\label{figure:spectra}
\end{figure*}

In Figure~\ref{figure:spectra}, the series of synthetic spectra are
presented for the selected epochs in the rest-frame as the observed spectra (S.\,Schulze, private communication; spectra are corrected for galactic extinction and the host in comparison to those publicly available): days~94, 231, 286, 362\footnote{The entire set of \textsc{artis} spectra between day~150 and day~850 after the explosion are available via {\textbf{\href{https://zenodo.org/doi/10.5281/zenodo.10473678}{https://doi.org/10.5281/zenodo.10473679}}}.}. We show three theoretical sets in the plots: (1) P250Ni25 is in red, (2) P250Ni34 is in blue, and (3) P250Ni34mix is in cyan. The observed spectra (shown in gray) are converted to the rest-frame\footnote{$F_\lambda^\mathrm{rest}= 4\pi F_\lambda^\mathrm{obs} \times (1+z) \times D^{\,2}$, where $z=0.166$ is redshift, and $D=882.6$\,Mpc is a distance to SN\,2018ibb.}. Each observed spectrum is smoothed with a median filter \citep[window size: ten data points, ][]{2007Image..book..Pratt} for demonstrative purpose. 
All spectra in Figure~\ref{figure:spectra} are normalised by a different factor for each subplot, which corresponds to a maximum flux value in a given subplot. We emphasise that synthetic light curves are computed with the \textsc{stella} code, and spectral synthesis is carried out with the \textsc{artis} code, therefore, some discrepancy can be seen between the flux in spectra at a particular epoch and magnitudes in broad bands. Nevertheless, \textsc{stella} and \textsc{artis} broad band light curves are consistent with each other.


Overall agreement between observed and synthetic spectra, particularly for our best fitting model P250Ni34, is good, specifically in redder wavelengths, although exact agreement is not fulfilled. At day~231 and later, there is a pronounced flux excess in blue wavelengths.
In fact, the difference between observed spectra and synthetic spectra are mainly seen below 5500\,\AA{}. The maximum in the residual ``spectra'' is located at the wavelength of 4000\,\AA{} and does not move noticeably between day~231 and day~362. The decreasing residual flux is likely explained by a decreasing size of the emitting region. This might be consistent with the findings by \citet{2024A&A...683A.223S} who claim that occultation of circumstellar matter (CSM) by the SN ejecta is present in SN\,2018ibb based on analysis of late time evolution of the [O\,II]\,$\lambda\lambda\,7318,\,7330$ lines.



The largest contribution to the flux comes from [Fe\,II] emission, but there are prominent lines from other ions. Below we mention a couple of prominent lines seen in the synthetic and observed spectra:
\begin{itemize}
\item{[Ca\,II]\,$\lambda\,3969$ is seen in modelled spectra at all epochs, however, it does not reproduce the observed line, partly because of additional flux in the blue wavelengths.}
\item{[O\,III]\,$\lambda\,5007$, which is formed in the low-density environment and indicates the presence of interaction, is not reproduced in the model spectra due to the low ionisation state of O (mostly neutral). }
\item{[O\,I]\,$\lambda\lambda\,6300,\,6364$ is present in both synthetic and observed spectra at day~231 and later, although the shape of the synthetic line differs from the observed one.}
\item{The blend [O\,II]\,$\lambda\lambda\,7318,\,7330$ and [Ca\,II]\,$\lambda\lambda\,7291,\,7323$ seen in SN\,2018ibb and is reproduced by our models to some extent. At earlier epoch (day~94) the unmixed models overestimate flux in this blend, while at later epochs (later than 2231~days) the models still display this blend, but the flux is underestimated. In the last shown epoch, day~362, the mixed model is able to reproduce the line quite well, which means that there should be some oxygen at low velocity and low electron density.}
\item{Synthetic line [O\,I]\,$\lambda\,7773$ can be considered to match the observed line through entire evolution, however, having slightly different strength and shape.}
\item{[Ca\,II]\,$\lambda\lambda\,8544,\,8664$ is well presented in synthetic spectra and explain this line seen in SN\,2018ibb.}
\end{itemize}
We also note that, spectra on day~600 are not nebular yet, since there are quite a few interactions between packets.


Presumably, the spectra of SN~2018ibb are a superposition of two types of spectra or two mechanisms, one of which is flux from the pure PISN explosion. In other words, there should be an additional source of energy input on top of the flux coming from the PISN itself, because the PISN models explain the light curves and partially spectra of SN~2018ibb considerably well, i.e. PISN radiation is a major contributor to the flux from SN\,2018ibb. 
The residual flux can be explained by interaction of SN ejecta with the surrounding CSM, which is consistent with \citet{2024A&A...683A.223S}. 

\subsection[The secondary energy source in SN\,2018ibb]{The secondary energy source in SN\,2018ibb}
\label{subsect:second}

SN\,2018ibb undergoes rebrightening, including variations, around day~50 after the maximum in $uvw2$, $uvm2$, $uvw1$, and $u$ bands and after day~200 in redder bands, which is not predicted by our models. A number of SLSNe also display bumps in their light curves, i.e. deviate from the linear decline \citep{2022ApJ...933...14H,2023ApJ...943...42C}. 
Rebrightening or bumps might be caused when the ejecta encounter clumps \citep[see e.g.,][]{2019A&A...629A..17D} or shells located at some distance from the progenitor \citep{2024MNRAS.527.11970}. \citet{2024A&A...683A.223S} revealed several signatures of CSM interaction  based on their analysis of spectra: (1) detection of an Mg II absorption line system moving at a significantly lower velocity than the SN ejecta; (2) simultaneous emergence of [O\,II] and [O\,III] emission at 30 rest-frame days after maximum; and (3) detection of the [O\,III] line being at 1000~days after peak in the rest-frame.

We use the bolometric light curve to analyse the luminosity excess between 42~day before the peak and 600~day after the peak. Based on the timing of the maximum (what means rise time), 120~days since the explosion for P250Ni34, the distance to CSM or a shell can be estimated via: $R_\mathrm{CSM}=t_\mathrm{max}\times v$\,, where $t_\mathrm{max}=120$~days, and $v=10^{\,10}$\,cm\,s$^{\,-1}$ is the velocity of the outer ejecta, what means that the distance to the shell is about $10^{\,17}$\,cm. 
Assuming the velocity of a CSM of about $3,000$\,km\,s$^{\,-1}$, as derived in \citet{2024A&A...683A.223S}, the matter had be expelled by the progenitor roughly 11~yrs 
before the PI explosion.
The luminosity excess integrated over the duration of bumps corresponds to radiated energy of $8.1\times10^{\,50}$~erg. 
If the ejecta collide with a shell or a blob, it means that the outermost part of the ejecta interacts with the blob, and not the entire ejecta. Assuming that kinetic energy of the outer ejecta is converted into radiated energy with the efficiency of, e.g., 90\,\%{}, i.e. $\mathrm{E}_\mathrm{rad}=\mathrm{E}_\mathrm{kin,out,ej}\times 90\,\%$,  the required mass of the ejecta interacting with the shell is about 0.02~\Msun{}.
Following the simplified approach from \citet{2018SSRv..214...59M}, the mass of the CSM equals 0.4~\Msun{}.
The detailed modelling of the ejecta interacting with the CSM and prediction of observational features, like properly calculated light curves and spectra, are beyond of the scope of the current paper and require a separate thorough study. 

If consider the interaction as a primary source, than the resulting light curve will not last that long, and a large amount of radioactive nickel will be needed to support late time luminosity. For instance, \citet{2017ApJ...835..266T} showed that the light curve of PTF12dam requires 43~\Msun{} of SN ejecta interacting with 37~\Msun{} of CSM and 6~\Msun{} of radioactive nickel. In their simulations the ``interaction'' part of the light curve lasts about 200~days and can explain the main maximum of PTF12dam, i.e. the interaction can presumably be the primary powering mechanism. However PTF12dam requires a secondary energy source to explain the later decline of the light curve. Nevertheless, we note that the light curve of PTF12dam is significantly narrower than that of SN\,2018ibb \citep[see Fig.~31, ][]{2024A&A...683A.223S}, and this scenario, specifically the interaction mechanism solely, is not relevant for SN\,2018ibb.

\vspace{3mm}
The next underlying question is: Which process stands behind the non-steady ejection of a stellar matter 11~yrs before an explosion?  In the case of massive and very massive stars, it corresponds to core carbon burning.
Single stellar evolution models, like our P250 model, experience massive wind mass-loss during hydrogen core burning, i.e. thousands of years before their PI explosion \citep{2013MNRAS.433.1114Y}, therefore, this matter being hydrogen-rich dissolves by the time of explosion. Moreover, SN\,2018ibb is hydrogen and helium-free, which means that the ejecta and CSM do not contain these elements.
\citet{2015A&A...573A..18M} suggest that pulsation-driven mass-loss in metal-free very massive stars may produce dense CSM, however, their findings are mostly applicable to SLSNe Type IIn.
We note that the amount of CSM required to explain the luminosity excess is very low, about 0.4~\Msun{} in total, as roughly estimated above. 
This amount of matter could not originate from previous binary interactions, such as a stellar merger or a common envelope ejection, which both would lead to a loss of about 10\,\% of a total mass of a system, i.e. much higher mass \citep[e.g.][]{2013MNRAS.434.3497G,2023LRCA....9....2R}. LBV-like eruptions caused by envelope inflation cannot be a mechanism for the production of H-free ejecta in our case \citep[see e.g.][]{2015A&A...580A..20S}. Moreover, a merger or a common envelope phase only years before the SN explosion would require extreme fine-tuning \citep{2012ApJ...752L...2C}.
Nevertheless, those very massive stars spend their late stages of evolution as He-stars. In the case of the P250 model, the final configuration is mainly a pure CO-core with a tenuous layer of helium, meaning that nuclear burning takes place near the surface. Any nuclear burning, waves excited by nuclear burning, helium shell burning \citep[c.f.\ Fig.~1 in][]{2017MNRAS.464.2854K} could cause ejection of negligible amount hydrogen-free material during the last years before the explosion \citep[e.g.,][and other studies]{2012MNRAS.423L..92Q,2016MNRAS.458.1214Q}. 

\subsection[ARTIS vs. SEDONA and SUMO]{$\mathrm{ARTIS}$ vs. $\mathrm{SEDONA}$ and $\mathrm{SUMO}$}

Spectral synthesis simulations for PISN models were also performed with \textsc{sedona} \citep{2011ApJ...734..102K}, \textsc{cmfgen} \citep{2013MNRAS.428.3227D}, and \textsc{sumo} \citep{2016MNRAS.455.3207J}. It is difficult to make a direct comparison because the mapped PISN models are different. The highest mass PISN models in \citet{2013MNRAS.428.3227D} are: a 100~\Msun{} He-star with 5~\Msun{} of {}$^{56}$Ni, which is too low to compare to our high mass model P250, and a relatively extended (146~\Rsun{}) H-rich blue supergiant model 210~\Msun{} with 21~\Msun{} of {}$^{56}$Ni, which cannot be used for comparison to our compact H-free P250 model. 

The input model for \textsc{sedona} and \textsc{sumo} simulations was a 130~\Msun{} helium-core model He130 \citep{2002ApJ...567..532H}. Our input model is a hydrogenic star with an initial mass of 250~\Msun{}, which loses all hydrogen during the earlier evolution and forms a He-core of 127~\Msun{}, what means it is close to the He130 model. Among differences are the total amount of radioactive nickel {}$^{56}$Ni produced during the PI explosion and amount of energy: 34~\Msun{} and 40~\Msun{}, and 82~foe{} and 87~foe in our P250 model and their He130 model, respectively. Therefore, we chose the model He130 used in \textsc{sedona} and \textsc{sumo} simulations and our model P250Ni34 to validate whether spectra for our P250Ni34 are in qualitative agreement with the published studies. 

\begin{figure}
\centering
\hspace{-5mm}\includegraphics[width=0.5\textwidth]{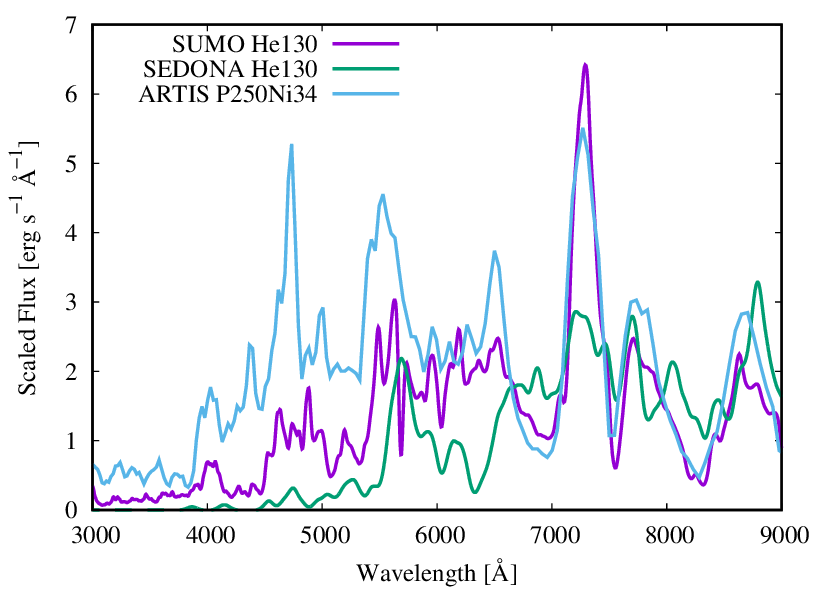}
\caption{\textsc{sumo} and \textsc{sedona} spectra for He130 model at day~400, and \textsc{artis} spectrum at day~371.}
\label{figure:comparison}
\end{figure}

In Figure~\ref{figure:comparison}, we show spectra of He130 model computed with \textsc{sedona} and \textsc{sumo} together with the spectrum for our model P250Ni34. 
The \textsc{sumo} and our \textsc{artis} spectra 
at corresponding epoch (250~day after the peak) have many common features. Among them are: [Si\,I]\,$\lambda\,4131$, [Ca\,II]\,$\lambda\lambda\,7291,\,7323$, [O\,I]\,$\lambda\,7773$, and many Fe\,II lines. The flux in \textsc{artis} spectra being higher than in \textsc{sumo} spectra can be explained by the fact that the ejecta at this epoch is still not fully transparent such as to produce a pure nebular emission.
Agreement between \textsc{sedona} and our \textsc{artis} spectra is poor, because \textsc{sedona} simulations are performed in the assumption of LTE approximation for ionisation states and level populations, while our \textsc{artis} simulations are carried out in full non-LTE.
The direct comparison of \textsc{artis} simulations in application to SN\,Ia spectra was done in separate studies, e.g., \citet{2020MNRAS.492.2029S} and \citet{2022A&A...668A.163B}. 

%% file: conclusions.tex

\section[Conclusions]{Conclusions}
\label{sect:conclusions}

In the present study, we examined a PISN origin of the
SLSN 2018ibb. We used the stellar evolution model of a very massive star with an initial ZAMS mass of 250~\Msun{} at metallicity $Z=0.001$ ($Z\cong 0.07\, Z_\odot${}, where $Z_\odot=0.014${}). The calculations were carried out using (1) the \textsc{genec} code to compute stellar evolution, (2) the \textsc{flash}
code to follow the dynamical phase of explosive oxygen burning, collapse and
PI explosion \citep{2017ApJ...846..100G,2017MNRAS.464.2854K}, and (3) the radiative-transfer code \textsc{stella} and \textsc{artis} to follow
up the hydrodynamical evolution of the ejecta and radiation field, and to
calculate observational properties, such as light curves and spectra. We found that the PI explosion of such a very massive star with 34~\Msun{} of radioactive nickel $^{56}$Ni provides a good match to bolometric and broad band light curves of SN 2018ibb during a long time interval, i.e. from the explosion and till 250 days after the peak.  
The synthetic spectra match the observed spectra to a large extent, including the appearance of many lines and the colour evolution. We emphasise that the best-fitting model is physically-consistent from the dawn of the progenitor life until its explosion becomes invisible, and the good match is exceptional given that our model has not been tuned to fit either light curves or spectra.

Nevertheless, our PISN models suffer from a deficit of a flux in the blue wavelengths between 3000\,\AA\ and 5000\,\AA{}, while the observed spectra clearly show such a significant flux. 
This indicates that there is a secondary source of energy powering the radiation field of SN\,2018ibb. Among the possibilities, we suggest interaction of the PISN ejecta with circumstellar material, which may produce the additional blue flux missing by the PISN model. Analysis of the light curve shows that a series of mass eruption happened before the final PI explosion. The amount of matter required to explain the flux excess is about 0.4~\Msun{}, which might originate from some activity during the latest phases of evolution of the star, caused by nuclear instabilities going on at the outermost atmosphere. We estimate that this material is located at a distance of 0.2\,pc and was ejected by a progenitor decades prior the PI explosion. The nature of the pre-explosion ejection might be instabilities caused by nuclear burning happening in the outer layers of the progenitor.